\documentclass[12pt]{aastex62}
\submitjournal{ApJ}

\pdfoutput=1

\shorttitle{GM\,Cep Variability}
\shortauthors{Huang et al. }

\begin{document}

%\hfill Today is \today

\title{ Diagnosing the Clumpy Protoplanetary Disk of the UXor Type Young Star GM Cephei
	}

\correspondingauthor{P.~C. Huang}
\email{pochiehhuang1@gmail.com}

\author{P.~C. Huang}
\affiliation{Graduate Institute of Astronomy, National Central University, 
300 Zhongda Road, Zhongli, Taoyuan 32001, Taiwan}

\author[0000-0003-0262-272X]{W.~P. Chen}
\affiliation{Graduate Institute of Astronomy, National Central University, 
300 Zhongda Road, Zhongli, Taoyuan 32001, Taiwan}
\affiliation{Department of Physics, National Central University, 
300 Zhongda Road, Zhongli, Taoyuan 32001, Taiwan}

\author{M. Mugrauer}
\affiliation{Astrophysikalisches Institut und Universit{\"a}ts-Sternwarte, 
	FSU Jena, Schillerg{\"a}$\beta$chen 2-3, D-07745 Jena, Germany}

% in alphabetical order 
	
\author{R. Bischoff}
\affiliation{Astrophysikalisches Institut und Universit{\"a}ts-Sternwarte, 
	FSU Jena, Schillerg{\"a}$\beta$chen 2-3, D-07745 Jena, Germany}

\author{J. Budaj}
\affiliation{Astronomical Institute, Slovak Academy of Sciences, 059 60 Tatransk\'a Lomnica, Slovak Republic}

\author{O. Burkhonov}
\affiliation{Ulugh Beg Astronomical Institute of the Uzbek Academy of Sciences, 
33 Astronomicheskaya str., Tashkent 100052, Uzbekistan}

\author{S. Ehgamberdiev}
\affiliation{Ulugh Beg Astronomical Institute of the Uzbek Academy of Sciences, 
33 Astronomicheskaya str., Tashkent 100052, Uzbekistan}

\author{R. Errmann}
\affiliation{Centre for Astrophysics Research (CAR), University of Hertfordshire, 
Hatfield Hertfordshire, AL10 9AB,UK}
\affiliation{Abbe Center of Photonics, Friedrich-Schiller-Universit{\"a}t Jena,
Max-Wien-Platz 1, D-07743 Jena, Germany}

\author{Z. Garai}
\affiliation{Astronomical Institute, Slovak Academy of Sciences, 059 60 Tatransk\'a Lomnica, Slovak Republic}

\author{H.~Y. Hsiao}
\affiliation{Graduate Institute of Astronomy, National Central University, 
300 Zhongda Road, Zhongli, Taoyuan 32001, Taiwan}

\author{C.~L. Hu}
\affiliation{Taipei Astronomical Museum, 363 Jihe Road, Shilin, Taipei 11160, Taiwan}

\author{R. Janulis}
\affiliation{Institute of Theoretical Physics and Astronomy, Vilnius University, Saul\.{e}tekio av. 3, 
10257 Vilnius, Lithuania}

\author[0000-0002-4625-7333]{E. L. N. Jensen}
\affiliation{Department of Physics and Astronomy, Swarthmore College, 500 College Avenue, Swarthmore, PA 19081, USA}

\author{S. Kiyota}
\affiliation{Variable Stars Observers League in Japan (VSOLJ), 7-1 Kitahatsutomi, Kamagaya, 
Chiba 273-0126, Japan}

\author[0000-0002-6757-8064]{K. Kuramoto}
\affiliation{Department of Cosmosciences, Hokkaido University, Kita 10, Nishi 8, Kita-ku, Sapporo, Hokkaido 060-0810, Japan}

\author{C.~S. Lin}
\affiliation{Graduate Institute of Astronomy, National Central University, 
300 Zhongda Road, Zhongli, Taoyuan 32001, Taiwan}

\author{H.~C. Lin}
\affiliation{Graduate Institute of Astronomy, National Central University, 
300 Zhongda Road, Zhongli, Taoyuan 32001, Taiwan}

\author{J.~Z. Liu}
\affiliation{National Astronomical Observatories, Xinjiang Observatory, Chinese Academy of Sciences, 
150, Science 1-Street, Urumqi, Xinjiang 830011, China}

\author{O. Lux}
\affiliation{Astrophysikalisches Institut und Universit{\"a}ts-Sternwarte, 
	FSU Jena, Schillerg{\"a}$\beta$chen 2-3, D-07745 Jena, Germany}

\author[0000-0001-9067-7653]{H. Naito}
\affiliation{Nayoro Observatory, 157-1 Nisshin, Nayoro, Hokkaido 096-0066, Japan}

\author{R. Neuh{\"a}user} 
\affiliation{Astrophysikalisches Institut und Universit{\"a}ts-Sternwarte, 
	FSU Jena, Schillerg{\"a}$\beta$chen 2-3, D-07745 Jena, Germany}

\author{J. Ohlert} 
\affiliation{University of Applied Sciences, Wilhelm-Leuschner-Strasse 13, D-61169 Friedberg, Germany}
\affiliation{Michael Adrian Observatory, Astronomie Stiftung Trebur, Fichtenstrasse 7, D-65468 Trebur, 
Germany}

\author[0000-0002-3326-2918]{E. Pak\v{s}tien\.{e}}
\affiliation{Institute of Theoretical Physics and Astronomy, Vilnius University, Saul\.etekio av. 3, 
10257 Vilnius, Lithuania}

\author{T. Pribulla}
\affiliation{Astronomical Institute, Slovak Academy of Sciences, 059 60 Tatransk\'a Lomnica, Slovak Republic}

\author{J.~K.~T. Qvam}
\affiliation{Department of Physics, University of Oslo, P.O. Box 1048 Blindern, NO-0316 Oslo, Norway}

\author{St. Raetz}
\affiliation{Freiburg Institute of Advanced Studies (FRIAS), University of Freiburg, Albertstra$\beta$e 19, 
D-79104 Freiburg, Germany}
\affiliation{Institute for Astronomy and Astrophysics T{\"u}bingen (IAAT), University of T{\"u}bingen, 
Sand 1, D-72076 T{\"u}bingen, Germany}

\author{S. Sato}
\affiliation{Astrophysics Department, Nagoya University, Nagoya, 464-8602, Japan}

\author{M. Schwartz}
\affiliation{Tenagra Observatory, 221 Calle Coco, Rio Rico, AZ 85648, USA}

\author{E. Semkov}
\affiliation{Institute of Astronomy and National Astronomical Observatory, 
Bulgarian Academy of Sciences, 72 Tsarigradsko Shosse Blvd., 1784 Sofia, Bulgaria}

\author{S. Takagi}
\affiliation{Department of Cosmosciences, Hokkaido University, Kita 10, Nishi 8, Kita-ku, Sapporo, 
Hokkaido 060-0810, Japan}

\author{D. Wagner} 
\affiliation{Astrophysikalisches Institut und Universit{\"a}ts-Sternwarte, 
	FSU Jena, Schillerg{\"a}$\beta$chen 2-3, D-07745 Jena, Germany}

\author[0000-0002-3656-4081]{M. Watanabe} 
\affiliation{Department of Applied Physics, Okayama University of Science 1-1 Ridai-cho, Kita-ku, Okayama, 
Okayama 700-0005, Japan}

\author{Yu Zhang}
\affiliation{National Astronomical Observatories, Xinjiang Observatory, Chinese Academy of Sciences, 
150, Science 1-Street, Urumqi, Xinjiang 830011, China}

\begin{abstract}
%r
UX Orionis stars (UXors) are Herbig Ae/Be or T Tauri stars exhibiting sporadic occultation  
of stellar light by circumstellar dust. GM\,Cephei is such a UXor in the young ($\sim4$~Myr) 
open cluster Trumpler\,37, showing prominent infrared excess, 
emission-line spectra, and flare activity. Our photometric monitoring (2008--2018) 
detects (1)~an $\sim$3.43~day period, likely arising from rotational modulation 
by surface starspots, (2)~sporadic brightening on time scales of days 
due to accretion, (3)~irregular minor flux drops due to circumstellar dust extinction, and 
(4)~major flux drops, each lasting for a couple of months with a recurrence time, 
though not exactly periodic, of about two years.  The star experiences normal reddening by 
large grains, i.e., redder when dimmer, but exhibits an unusual ``blueing" phenomenon in that the star turns blue 
near brightness minima.  The maximum extinction during relatively short 
(lasting $\leq 50$~days) events, is proportional to 
the duration, a consequence of varying clump sizes.  For longer events, the extinction is independent 
of duration, suggestive of a transverse string distribution of clumps.  
Polarization monitoring indicates an optical polarization varying $\sim3\%$--8$\%$, with the 
level anticorrelated with the slow brightness change.  
Temporal variation of the unpolarized and polarized light sets constraints on the size and orbital distance of the 
circumstellar clumps in the interplay with the young star and scattering envelope.  These transiting 
clumps are edge-on manifestations of the ring- or spiral-like structures found recently in young 
stars with imaging in infrared of scattered light, or in submillimeter of thermalized dust emission. 
\end{abstract}

\keywords{ circumstellar matter --- occultations --- protoplanetary disks ---  stars: individual (GM~Cephei) --- 
stars: pre-main sequence --- stars: variables: T Tauri, Herbig Ae/Be  
         }

%sec1
\section{Introduction} \label{sec:intro}

Circumstellar environments are constantly changing.  A young stellar object (YSO), with 
prominent chromospheric and coronal activities, interacts intensely with the surrounding 
accretion disk by stellar/disk winds and outflows.  The first few million years of 
the pre-main-sequence (PMS) evolution coincide with the epoch of possible planet formation, 
during which grain growth, already taking place in prestellar molecular cores up to micron sizes, 
continues on to centimeter sizes, and then to planetesimals \citep{nat07}.  
The detailed mechanism to accumulate planetesimals and to eventual planets is still uncertain.  
Competing theories include planetesimal 
accretion \citep{wei00} versus gravitational instability \citep{saf72,gol73,joh07}.  Given the 
ubiquity of exoplanets, planet formation must be efficient to complete with the 
dissipation of PMS optically thick disks in less than 10~Myr \citep{mam04,bri07,hil08}.

YSOs are known to vary in brightness.  Outbursts arising from intermittent mass accretion events are 
categorized into two major classes: (1)~FU Ori-type stars (or FUors) showing erupt brightening up to 6~mag 
from quiescent to the high state in weeks to months, followed by a slow decline in decades 
\citep{har85}, and (2)~EX Lup-type stars (EXors) showing brightening up to 5~mag, sometimes recurrent, 
with roughly the same timescale of months in both rising and fading \citep{her89}.  Sunlike PMS 
objects, i.e., T~Tauri stars, may also display moderate variations in brightness and colors \citep{her94} 
due to rotational modulation by magnetic/chromospheric cool spots or accretion/shocking hot spots 
on the surface. There is an additional class, owing its variability to extrinsic origin, of UX\,Ori 
type stars \citep[UXors;][]{her94}, that displays irregular dimming caused by circumstellar dust 
extinction.   In addition to the prototype UX\,Ori itself, examples of UXors include CO\,Ori, RR\,Tau, 
and VV\,Ser. 

The YSO dimming events can be further categorized according to the levels of extinction and the 
timescales.  The ``dippers'' \citep{cod10}, with AA\,Tau being the prototype \citep{bou99, bou03}, 
have short (1--5 days) and quasi-periodic events thought to originate from occultation 
by warps \citep{ter00,cod14} or by funnel flows \citep{bli16} near the disk truncation radius and induced by 
the interaction between the stellar magnetosphere and the inner disk \citep{rom13}.  The ``faders,'' 
with KH\,15D being the prototype \citep{kea98,ham01}, show prolonged fading events, each lasting for months to 
years with typically large extinction up to several magnitudes, thought to be caused by occultation by 
the outer part of the disk \citep{bou13,rod15,rod16}.  The target of this work, GM\,Cephei 
(hereafter GM\,Cep), a UXor star known to have a clumpy dusty disk \citep{che12}, displays both dipper and 
fader events. 

As a member of Trumpler (Tr)~37, a young (1--4~Myr, \citep{mar90,pat95,sic05,err13}) star cluster 
as a part of the Cepheus OB2 association, GM\,Cep (R.A.=21$^{\rm h}$38$^{\rm m}$17$\fs32$, 
Decl.=+57\degr31\arcmin22\arcsec, J2000) possesses observational properties typical of a T~Tauri star, 
such as emission spectra, infrared excess, and X-ray emission \citep{sic08,mer09}.  
{\it Gaia}/DR2 \citep{bro18} measured a parallax of $\varpi=1.21\pm0.02$~mas ($d=826_{-13}^{+14}$~pc), 
consistent with being a member of Tr\,37 at $\sim870$~pc \citep{con02}.

The spectral type of GM\,Cep reported in the literature ranges from a late F \citep{hua13} 
to a late G or early K \citep{sic08}.  The star has been measured to have a disk accretion rate up to 
$10^{-6} M_\sun$~yr$^{-1}$, which is thought to be 2--3 orders higher than the median 
value of the YSOs in Tr\,37 and is 1--2 orders higher than those of typical T Tauri 
stars \citep{gul98,sic08}.  The broad spectral lines suggest a rotation 
$ v \sin i\sim43.2$~km~s$^{-1}$ much faster than the average $v\sin i\sim10.2$~km~s$^{-1}$ 
of the members of Tr\,37 \citep{sic08}.  

\citet{sic08} presented a comprehensive collection of data on GM\,Cep, including optical/infrared photometry 
and spectroscopy, plus millimeter line and continuum observations, along with the young stellar population in 
the cluster Tr\,37 and the Cep OB2 association \citep[See also][]{sic04,sic05,sic06a,sic06b}.  
Limited by the time span of their light curve, \citet{sic08} made the incorrect 
conclusion that the star belonged to the EXor type.  Later, with a century-long light curve derived 
from archival photographic plates, covering 1895 to 1993, \citet{xia10} classified 
the star as a UXor, which was confirmed by subsequent intense photometric monitoring 
\citep{che12,sem12,sem15,hua18}.  
\citet{che12} speculated on a possible recurrent time of $\sim1$ yr based on a few major 
brightness dimming events, but this was not substantiated by \citet{sem15}.

GM\,Cep has been studied as a part of the Young Exoplanet Transit Initiative (YETI) project \citep{neu11}, which 
combines a network of small telescopes in distributed time zones to monitor young star clusters, 
with the goal to find possible transiting exoplanets \citep{neu11}.  
Any exoplanets thus identified would have been newly formed or in the earliest evolution, providing a 
comparative sample with the currently known exoplanets that are almost exclusively found in the 
general Galactic fields, so are generally older).  While so far YETI has detected only exoplanet 
candidates \citep{gar16,rae16}, 
the data set serves as a valuable inventory for studies such as stellar variability \citep{err13,fri16}.

The work reported here includes light curves in $BVR$ bands on the basis of the photometry collected 
from 2008 to 2018.  Moreover, polarization measurements in $g^{\prime}$-, $r^{\prime}$-, and $i^{\prime}$-bands 
have been taken at different brightness phases, enabling simultaneous photometric and polarimetric diagnosis of 
the properties of the circumstellar dust clumps that cause the UXor variability.
\S\ref{sec:data} summarizes the data used in this study, including those collected in the 
literature, and our own photometric and polarimetric observations.  \S\ref{sec:results} 
presents the results of photometric, color, and polarimetric variations.  On the temporal 
behavior of these measurements, we then discuss in \S\ref{sec:disk} the implications on 
the properties of the dust clumps around GM\,Cep.  We summarize our findings in 
\S\ref{sec:conclusion}.

%sec2
\section{Data Sources and Observations} \label{sec:data}

Optical data of GM\,Cep consist mostly of our own imaging photometry since mid-2008, and polarimetry 
since mid-2014, up to mid-2018.  These are supplemented by data adopted from the American Association 
of Variable Star Observers (AAVSO) database, covering timescales from days/weeks to years.  
\citet{sic08} summarized the photometry from the literature, e.g., those of \citet{mor39} \citet{suy75}, and 
\citet{kun86}, and from databases such as VizieR, SIMBAD, and SuperCOSMOS \citep{mon03}, along with the 
infrared data from $IRAS$ and MSX6C. \citet{xia10} expanded the light-curve baseline and presented 
a-century-long photometric measurements,
with a photometric uncertainty of $\sim0.15$~mag, derived from the photographic plates collected at the Harvard 
College Observatory and from Sonneberg Observatory.  Previous optical monitoring data include 
those reported by \citet[][in BVR covering end of 2009 to 2011]{che12}, by \citet{sem12}, and 
by \citet[][in $UBVRI$ to end of 2014]{sem15}.  The AAVSO data were adopted only from the observer ``MJB'' 
after checking photometric consistency with our results. 

%2.1
\subsection{Optical photometry}

The imaging photometry covering 10 years has been acquired by 16 telescopes, including seven of the 
YETI telescopes \citep{neu11}.  The Tenagra Observatory in Arizona and Lulin Observatory in Taiwan 
contributed about four-year baseline coverage each from mid-2010 to mid-2018, respectively.  The 
Tenagra~II telescope, a 0.81m, Ritchey--Chr\'etien type telescope, carried out the $BVR$ monitoring 
from 2010 October to 2014 June. No observations were taken in July/August because of the monsoon season, 
or during February/March because of the invisibility of the target.  The SLT 0.4 m telescope, 
located at Lulin Observatory, acquired a few data points in $BVR$ bands every night from 2014 September
to date, weather permitting. Technical parameters of additional telescopes contributing 
to the data are listed in Table~\ref{tab:tele}.

%In addition to the Tenagra and SLT telescopes, some data were taken by the 1-m LOT of Lulin Observatory, the 0.25-m CTK-II in Jena, 0.6~m in Star{\'a} Lesn{\"a} Observatory, the 
%1.2-m T1T in Michael Adrian Observatory, the 0.61~m of Swarthmore College, the 
%1.5~m in Maidanak Observatory, the 1~m ESA's Optical Ground Station (OGS) at the 
%Observatoris del Teide on Tenerife, 1.5~m in Palomar Observatory, the 1.6~m Pirka telescope of Hokkaido University in Nayoro, and the Schmidt 0.6-m STK in Jena. The telescope 
%parameters are listed in Table~\ref{tab:tele}. 

For each observing session, darks and bias frames were obtained every night when science 
frames were taken, except for the STK and CTK-II, for which darks already include biases. The 
sky flats were obtained when possible. For those nights without sky flats, we used the flats 
from the nearest previous night. The standard reduction with dark, bias, and flat field correction 
was performed with IRAF. For the Maidanak Observatory, Nayoro Observatory, and the ESA's
OGS, the images were only corrected with bias and flat because of the low temperatures of the CCD 
detectors used.  

The brightness of GM\,Cep and photometric reference stars was each measured with the aperture photometry 
procedure ``aper.pro'' of IDL, which is similar to the ``IRAF/Daophot'' task, with an aperture radius 
of 8\farcs5 for the target, and an annulus of the inner radius of 9\farcs5 and outer radius of 13\arcsec 
for the sky.  The seven reference stars from \citet[][their Table~2]{xia10} 
were originally used by \citet{che12}, but later we found that Star~A varied at $\sim0.1$~mag 
level, and Star~E was likely a member of the young cluster, so would be likely also variable. 
Excluding these two stars, the remaining five were used as reference stars in the differential 
photometry of GM\,Cep reported here.  

Photometric measurements at multiple bands were taken at different epochs in a night, and 
sometimes with different telescopes.  In order to facilitate a quantitative comparison, e.g., 
between the $B$- and $V$-band light curves, and hence the $B-V$ color curve, the epoch 
of each observation was rounded to the nearest integer Modified Julian Date (MJD), 
and the average in each band 
was taken within the same MJD.  For periodicity analysis, the actual timing was used, so 
there would be no round-off error.

%%%%%%%%%%%%%%%%%%%%%%%%%%%
\begin{deluxetable}{lllcl cc}
\tablecaption{ Parameters of Telescopes  
		}
\tabletypesize{\scriptsize}
\tablehead{
	\colhead{Observatory/Telescope} & \colhead{CCD Type} & \colhead{Size (pixels)} & \colhead{Pixel Size ($\mu$m)} & 
	\colhead{FOV (arcmin$^2$)} & \colhead{RON ($e^{-}$)} & \colhead{\# Nights}
	  }
\startdata
\multicolumn{6}{c}{ YETI Telescopes  } \\  \hline
0.4~m SLT (Lulin)  & E2V~42-40 & 2048$\times$2048 & 13.5 & 30.0$\times$30.0 & 7 & 541\\
0.81~m TenagraII (Tenagra) & SITe~SI-03xA & 1024$\times$1024 & 24 & 14.8$\times$14.8 & 29  & 463\\
0.25~m CTK-II (Jena)$^{a}$ & E2V~PI47-10 & 1056$\times$1027 & 13 & 21.0$\times$20.4 & 7 & 104\\
0.6~m STK (Jena)$^{b}$ & E2V~42-10 & 2048$\times$2048 & 13.5 & 52.8$\times$52.8 & 8  & 79\\
1.0~m LOT (Lulin) &  Apogee~U42 & 2048$\times$2048 & 13.5 & 11.0$\times$11.0 & 12  & 48\\
0.61~m RC (Van de camp) & Apogee~U16M & 4096$\times$4096 & 9 & 26.0$\times$26.0 & 7  & 13\\
0.6~m Zeiss 600/7500 (Stara Lesna) & FLI~ML~3041 & 2048$\times$2048 & 15 & 14.0$\times$14.0 & 5  & 11\\  \hline
\multicolumn{6}{c}{ Other Telescopes } \\  \hline
1.6~m Pirka (Nayoro)$^{c}$ &  EMCCD~C9100-13 & 512$\times$512 & 16 & 3.3$\times$3.3 & 13  & 133\\
1.5~m AZT-22 (Maidanak) & SI~600~Series & 4096$\times$4096 & 15 & 16.0$\times$16.0 & 5  & 120\\
1.0~m NOWT (XinJiang) & E2V~203-82 & 4096$\times$4096 & 12 & 78.0$\times$78.0 & 5  & 108\\
1.2~m T1T (Michael Adrian)  & SBIG~STL-6303 & 3072$\times$2048 & 9 & 10.0$\times$6.7 & 15  & 12\\
0.51~m CDK (Mayhill)   &   FLI~ProLine~PL11002M & 4008$\times$2072 & 9 & 36.2$\times$54.3 & 9  & 12\\
1.0~m ESA's OGS (Teide)$^{d}$  & Roper~Spec~Camera & 2048$\times$2048 & 13.5 & 13.76$\times$13.76 & 8 & 10\\
1.5~m P60 (Palomar) & AR-Coated~Tektronix & 2048$\times$2048 & 24 & 11.0$\times$11.0 & 9  & 7\\
0.35~m ACT-452 (MAO) & QSI~516 & 1552$\times$1032 & 9 & 37.6$\times$25.0 & 15 & 2\\
\enddata
\label{tab:tele}
\tablecomments{
$^{a}$\citet{mug16}; 
$^{b}$\citet{mug10}; 
$^{c}$Nayoro observatory equips EMCCD camera with their Multi-Spectral Imager (MSI) instrument\citep{wat12};
$^{d}$\citet{sch14}.
}
\end{deluxetable} 
%%%%%%%%%%%%%%%%%%%%%

%%%%%%%%%%%%%%%%%%%%%
\begin{deluxetable}{cccccc}
\tablecaption{Photometric reference stars adopted from \citet{xia10} \label{tab:ref}  }
\tablehead{
	\colhead{Refe Star} & \colhead{R.A. (J2000) (deg)} & \colhead{Decl. (J2000) (deg)} & 
	\colhead{ $B$ (mag)} & \colhead{ $V$ (mag)} & \colhead{ $R$ (mag)}  
	  }
\startdata
Star B & 324.529226 & 57.508117 & 16.015 & 14.961 & 14.364 \\
Star C & 324.563184 & 57.492816 & 15.445 & 14.837 & 14.455 \\
Star D & 324.543391 & 57.505287 & 15.333 & 14.357 & 13.770 \\
Star F & 324.586443 & 57.487231 & 14.389 & 13.358 & 12.770 \\
Star G & 324.600939 & 57.556202 & 13.374 & 12.829 & 12.513 \\
\enddata
\end{deluxetable} 
%%%%%%%%%%%%%%%%%%%%%%%%%%%%%%%

\subsection{Optical Polarimetry}

The optical polarization of GM\,Cep was measured by TRIPOL2, the second unit of the Triple-Range 
Imaging POLarimeter (TRIPOL, Chen et al. 2019, in preparation) attached to the LOT. This imaging polarimeter measures 
polarization in the Sloan $g^{\prime}$-, 
$r^{\prime}$-, and $i^{\prime}$-bands simultaneously by rotating a half-wave plate to four angles, 
0\degr, 45\degr, 22\degr.5, and 67\degr.5.  To reduce the influence by sky conditions, every polarization 
measurement reported in this work was the mean value of at least fuve sets of images having nearly the same 
counts in each angle.  This compromises the possibility to detect polarization variations on 
timescales of less than about an hour, but ensures the reliability of nightly measurements.  

For TRIPOL2, we acquired the sky flats if weather allowed, or else we used the sky flats from 
the nearest adjacent night. Several unpolarized and polarized standard stars \citep{sch92} 
were observed to calibrate the instrumental polarization and angle offset (Chen et al. 2019, in preparation). 
The correction for the dark and flat field was performed for all the images 
following the standard reduction procedure. The fluxes at four angles were measured 
with aperture photometry, and the Stokes parameters ($I$, $Q$, and $U$) 
were then calculated, from which the polarization percentage ($P=\sqrt{Q^2 + U^2}/I$) 
and position angle ($\theta=0.5\arctan(U/Q)^{-1}$) were derived.  A typical accuracy 
$\Delta P \lesssim 0.3$\% in polarization could be achieved in a photometric night 
(Chen et al. 2019, in preparation).  

%sec3
\section{Results and Discussions} \label{sec:results}

\subsection{Photometric Variations}

\begin{figure}[htb!]
\centering 
\includegraphics[angle=0, width=1.0\textwidth]{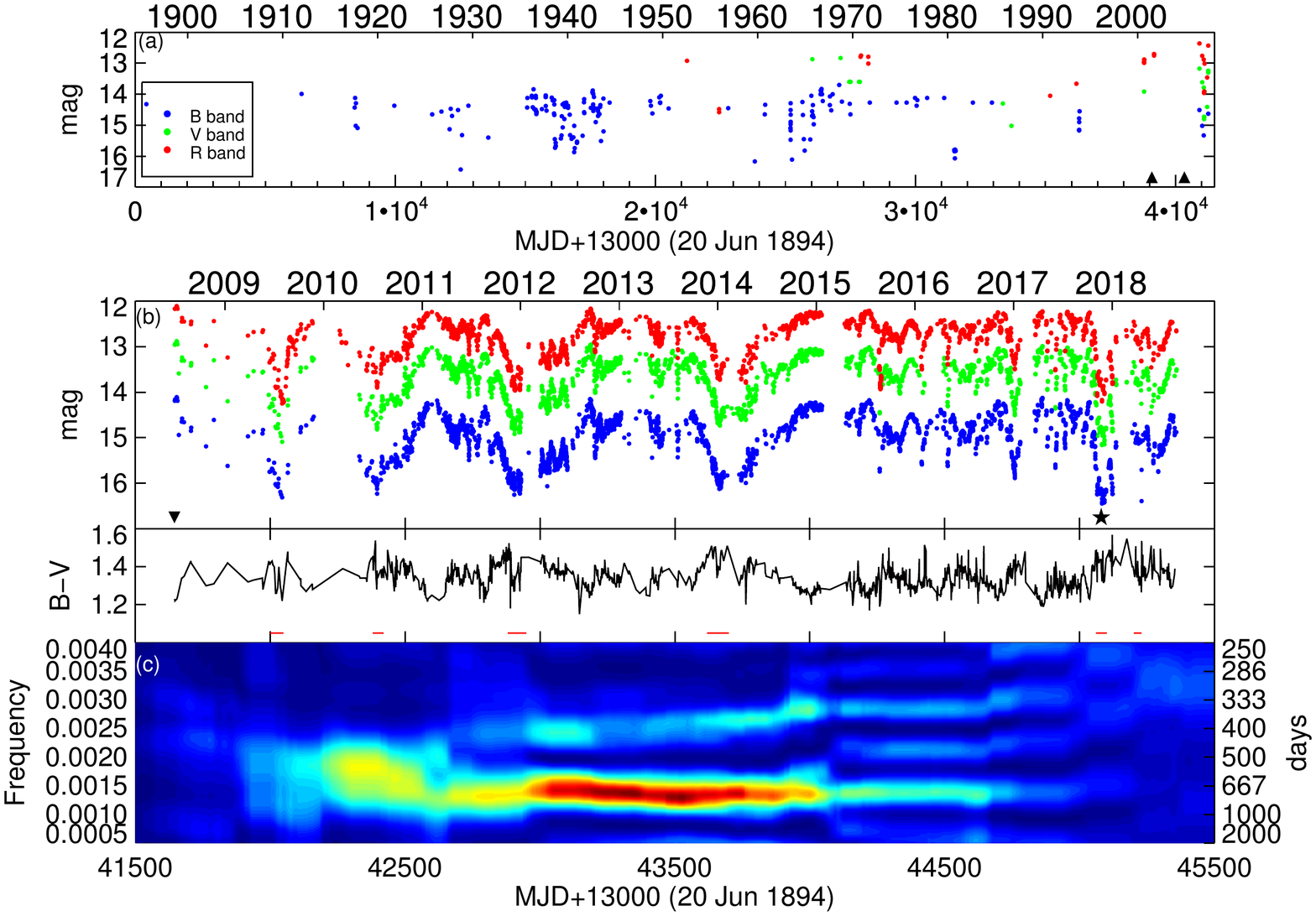}
\caption{The light curve of GM\,Cep from 1894 to 2018. 
    (a)~The century-long data reported by \citet{sic08} and \citet{xia10}.  
	(b)~The light curves and $(B-V)$ color curve from 2008 to 2018 reported in this work.
    Epochs at which spectral measurements were reported in the literature are marked, with a triangle
	symbol for \citet{sic08}, an upside down triangle for \citet{sem15} and an asterisk for \citet{gia18}. 
	(c)~Dynamical period analysis of the input light curve of (b), with a 
	window size of 2000 days and a step of 1 day.  The color represents the power 
	of the periodogram, from high in red to blue.  The vertical axis represents either the 
	frequency (on the left) or the corresponding period (right).
	}
\label{fig:lc}
\end{figure}

Figure~\ref{fig:lc} exhibits the light curves of GM\,Cep, including data taken from the literature 
covering more than a century since 1895 (Figure~\ref{fig:lc}a), and our intense multiband 
observations starting in 2008 (Figure~\ref{fig:lc}b).     
Since last reported \citep{che12,sem12,sem15}, the star continued to show abrupt brightness changes.  
There are three main kinds of variations.  Most noticeable are the major flux drops, 
$\sim1$--$2.5$~mag at all $B$-, $V$-, and $R$-bands, with prominent ones, each lasting for 
months, occurring in mid-2009, mid-2010, 2011/2012, beginning of 2014, 
end of 2016, and end of 2017 \citep{mun17}.  The list is not complete, limited
by the time coverage of our observations.  In addition, there are minor flux drops ($\sim0.2$--1~mag), each 
with the duration of days to weeks.  The third kind, with a typical 
depth of $0.05$~mag and occurring in a few days, is not discernible on the display 
scale of Figure~\ref{fig:lc}, and will be discussed later.

%fig1

\subsubsection{Periodicity Analysis}

\noindent{\it Deep Flux Drops}

\begin{figure}[htb!]
\centering
\includegraphics[angle=0,width=0.8\textwidth]{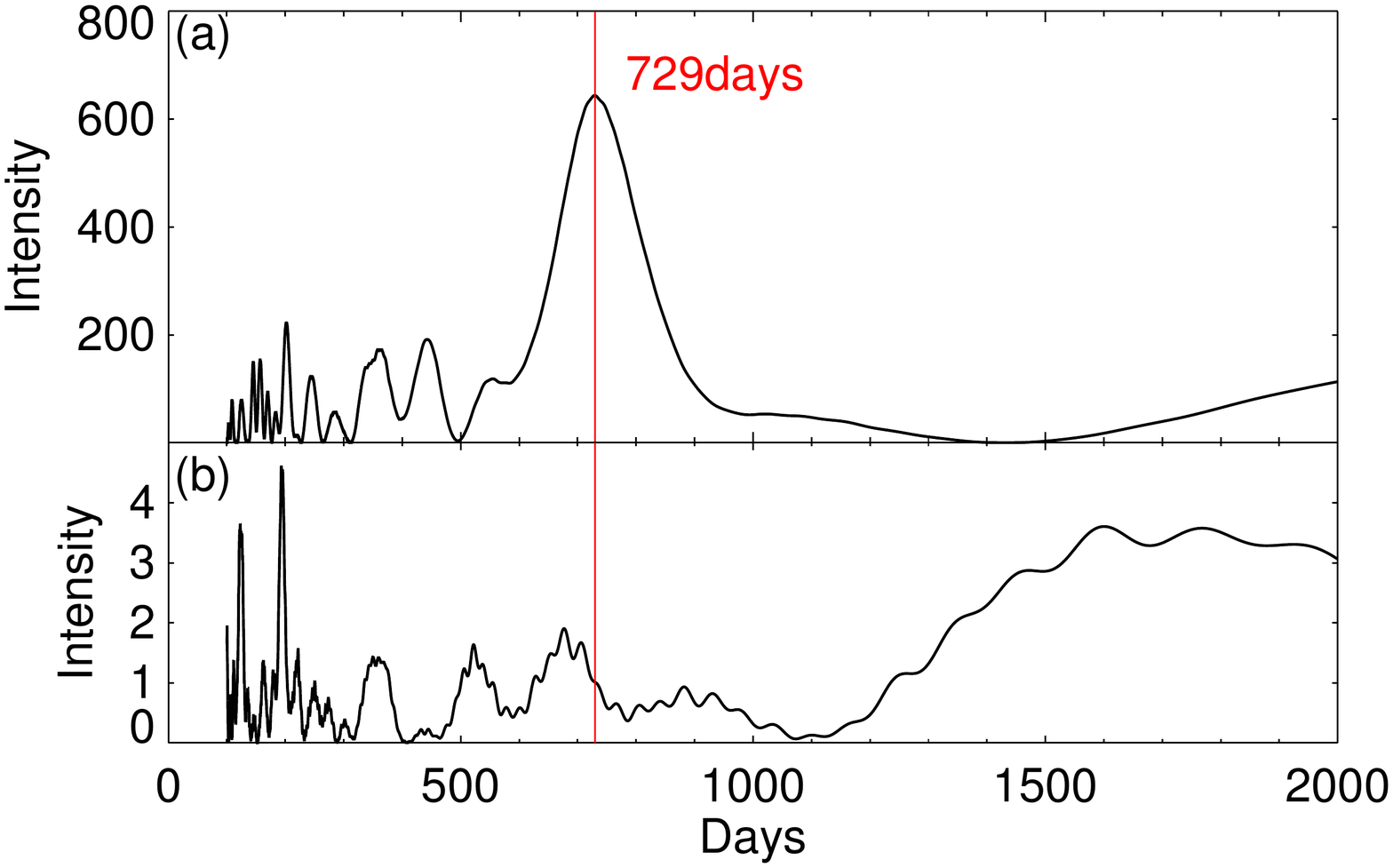}
	\caption{(a)~The periodogram of the V-band light curve,  where the red 
	line marks the peak of the power spectrum. 
	(b)~The periodogram of the sampling function.}
\label{fig:period}
\end{figure}

The UXors are thought to have irregular extinction events, despite 
the attempts to search for cyclic variability \citep{gri98,ros99}.
For GM\,Cep, period analysis by the Lomb--Scargle algorithm \citep{lom76,sca82} was performed, and the result 
is shown in Figure~\ref{fig:period}.  A significant power is seen at $\sim730$ days, which does not show up 
in the power spectrum of the sampling function (i.e., a constant magnitude at each sampling point).  
The secondary peak around 350~days, also visible in the sampling function, is the consequence of 
annual observing gaps.  
A dynamical period analysis was performed by repetitive Lomb--Scargle computation within a running window
of 2,000 days with a moving step of one day. For example, the power spectrum at date 42500 (plus MJD$+$13000) was 
calculated by the data within the window ranging from 41500 to 43500. Enough padding was applied to the edges of the 
light curve. A peak around $\sim700$~days persists, evidenced in Figure~\ref{fig:lc}c.	

An independent investigation of the periodicity was performed by computing the autocorrelation 
function.  The light curve was resampled to be equally spaced with a step of one day, and for 
each day, the average of data within 300 days from date 41500 to date 45500, or within 100 days 
from date 42300 to date 45500, centered on the day was adopted.  A time lag of $\sim700-800$ days 
is reaffirmed.   This is the timescale between the few prominent minima (i.e., near 42900 and 43700).

%fig2

\noindent{\it Rotational Modulation} 

\begin{figure}[htb!]
\centering
\includegraphics[width=0.8\textwidth]{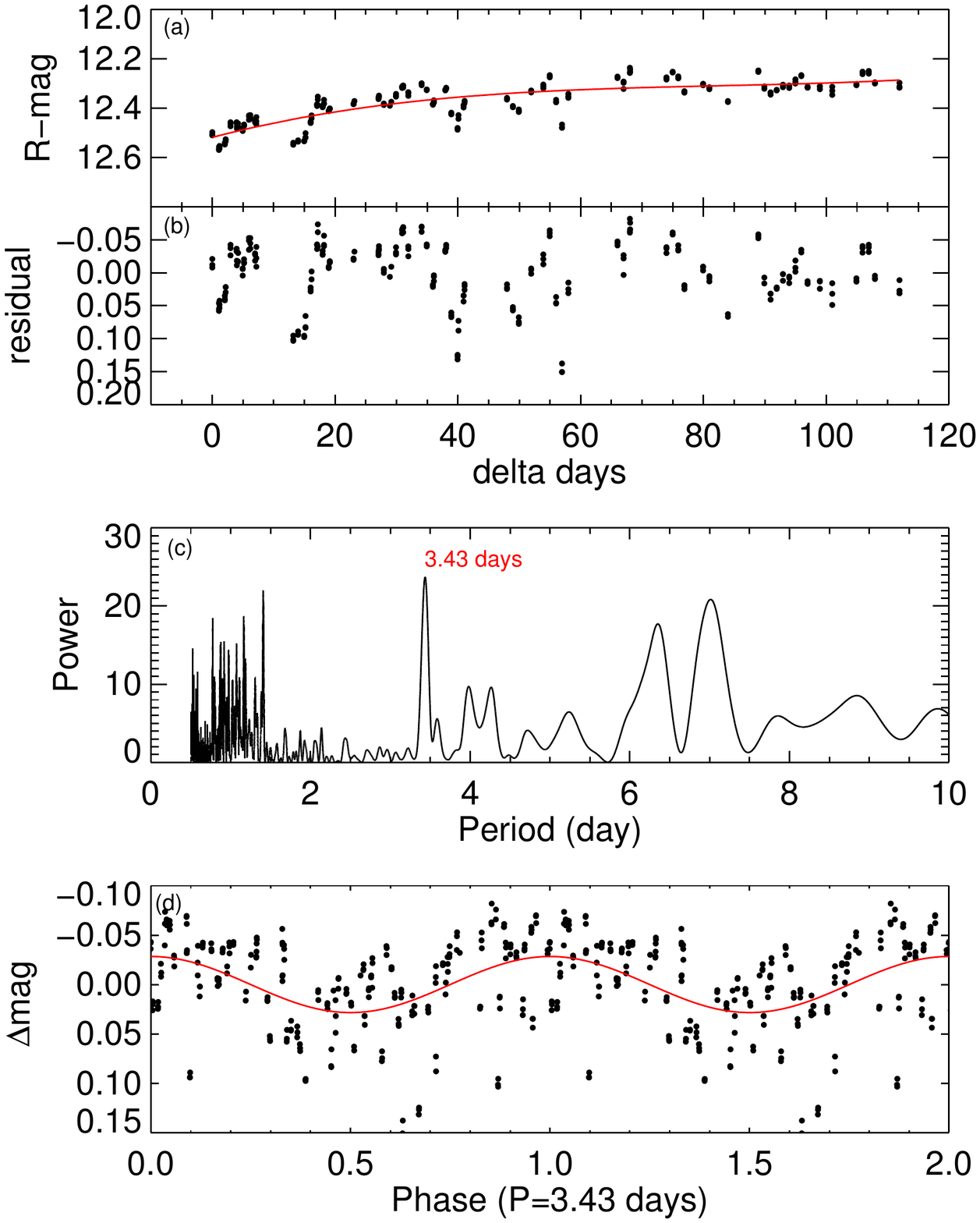}
\caption{(a)~The bright state in mid-2014 of the $R$-band light curve. 
     (b)~The scaled light curve after removal of the slow-varying trend.  
    (c)~The power spectrum of (b), from which a period of 3.43~days is detected.
	(d)~The folded light curve with $P=3.43$~days found in (c). The solid 
	curve shows the best-fit sinusoidal function. }
\label{fig:spots}
\end{figure}

To investigate possible variability on much shorter time scales, we extracted the segment of the 
light curve from mid-2014 to the end of 2014, when the star was in the bright state  
so that there should be little influence by major flux drops.  The light curve was fitted with, and 
then subtracted by, a third-order polynomial function to remove the slow-varying trend. 
The Lomb--Scargle analysis led to an identification of a period of $\sim3.43$ days in the 
detrended light curve, and Figure~\ref{fig:spots} exhibits the original and the detrended light curves, 
together with the power spectrum and the folded light curve.  This variation is caused 
by modulation of stellar brightness by dark spots on the surface with the rotational period of the star
\citep{str09}.  Note that this period 
coincides roughly with the expected rotational period of a few days for the star, given its measured 
rotation $ v \sin i\sim43$~km~s$^{-1}$, and a radius of a few solar radii, estimated from the 
PMS evolutionary tracks \citep{sic08}.

\begin{figure}[htb!]
\centering
 \includegraphics[angle=0,width=\textwidth]{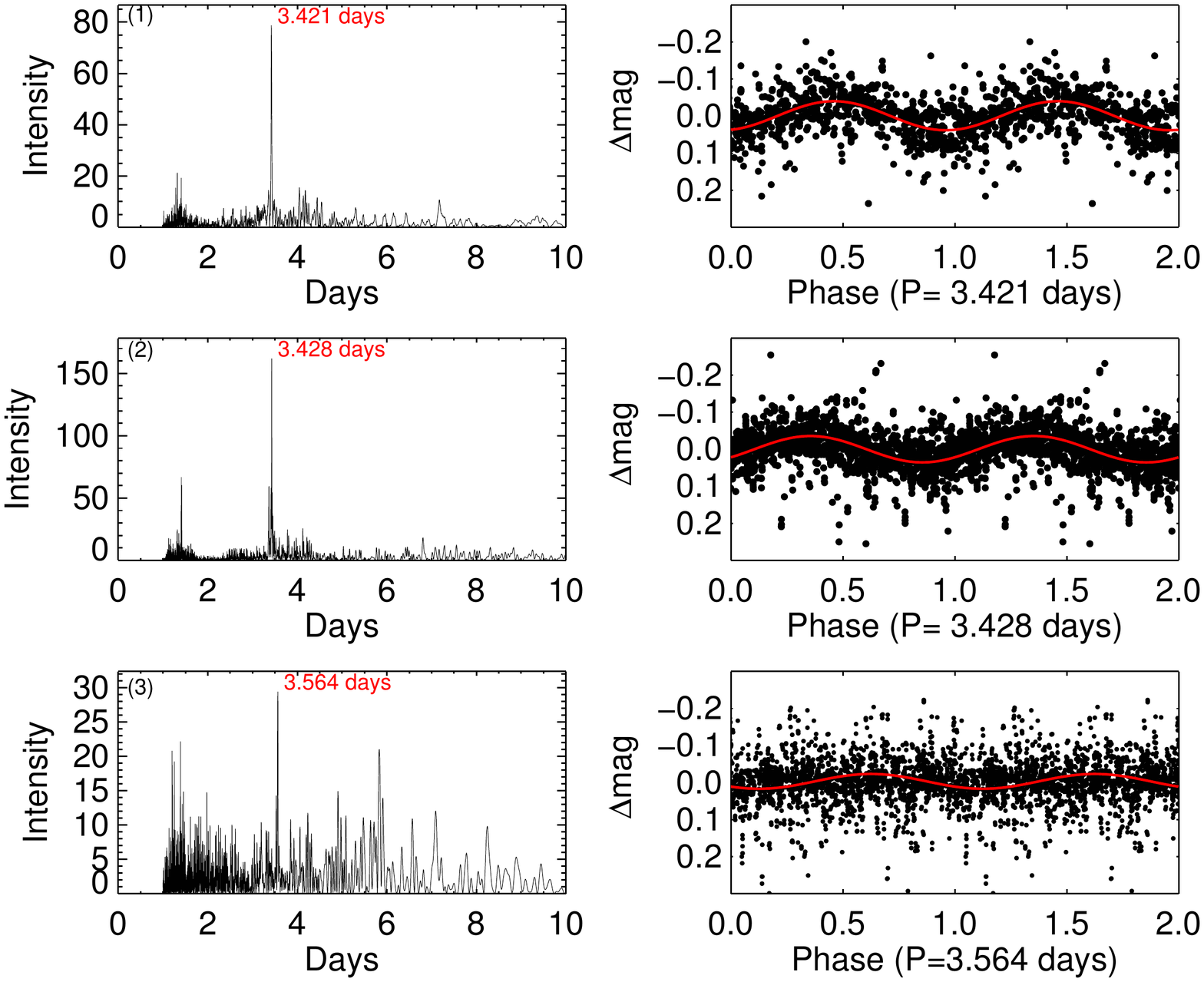}
	\caption{Power spectrum and phased light curve for (plus MJD+13000) 
         (a) 41500--43000, (b) 43000--44250, and (3) 44250--45500.  In each 
	 case the solid curve is the best-fit sinusoidal function, from which 
	 the amplitude is derived.}
\label{fig:segments}
\end{figure}

Guided by the periodicity derived from the short segment of the light curve, we then processed 
the entire light curve using a more aggressive detrend technique than a polynomial fit 
to deal with the large fluctuations.  The original light curve was 
smoothed by a running average, with an eight-day window.  This effectively removes low-frequency 
signals slower than about 10~days.  To investigate possible period changes, we divided the light 
curve into three segments, with the MJD ranges (plus MJD+13000) (1)~41500 to 43000, 
(2)~43000 to 44250, and (3)~44250 to 45500, respectively, based on a judicious choice to have 
sufficiently long trains of undersampled data to recover periods on time scales of days.  
Figure~\ref{fig:segments} presents the power spectrum and the phased light curve for each segment, 
and in each case a significant period stands out, with the period and amplitude, $P_1=3.421$~days, 
$A_1=0.039$~mag $P_2=3.428$~days, $A_2=0.036$~mag, and $P_3=3.564$~days, $A_3=0.020$~mag.  
The seemingly large scattering in each folded light curve is not the noise in the data, but  
the intrinsic variation in the star's brightness, e.g., by differing total starspot 
areas.  Because such a variation is not Gaussian, 
a least-squares analysis may not be appropriate to render a reliable estimate of the 
amplitude.  Still, the sinusoidal behavior seems assured.

Therefore, a rotation period of roughly 3.43~days is found to persist throughout 
the entire time of our observations.  Moreover there is marginal evidence of a 
lengthening period with a reduction in amplitude.  This can be understood 
as latitudinal dependence of the occurrence of starspots due to surface differential 
rotation, in analog to the solar magnetic Schwabe cycle, in which sunspots first appear in heliographic 
mid-latitudes, and progressively more new sunspots turn up (hence covering 
a larger total surface) toward the equator (hence with shorting rotational periods).  
GM\,Cep therefore has an opposite temporal behavior, suggestive of an alternative 
dynamo mechanism at work \citep[e.g.,][]{kuk11}.  Further observations 
with a shorter cadence should be able to confirm this period shift and to provide a 
more quantitative diagnostic.      

%%%%%%%%%%%%%%%%
%fig3
%%%%%%%%%%%%%%%%

%%%%%%%%%%%%%%%%
%fig4
%%%%%%%%%%%%%%%%

The detrended light curve shows mostly dimming events with occasional brightening episodes.  
The dimming must be the consequence of rotational modulation by surface starspots, whereas the brightening 
arises from sporadic accretion. The amplitude $\lesssim0.2$~mag is consistent with the 0.01--0.5~mag 
variation range typically observed in T Tauri stars caused by cool or hot starspots \citep{her94}.  Also, the 
amplitude of variation is marginally larger at shorter wavelengths, namely in $V$ and $B$, lending 
evidence of accretion.  

The excessive accretion rate of GM\,Cep reported by \citet{sic08}, $10^{-7}$ to 
$5 \times 10^{-6} M_\sun$~yr$^{-1}$, was estimated by the $U$-band luminosity \citep{gul98}.  
Using the H$_\alpha$ velocity as an alternative diagnostic tool \citep{nat04}, the accretion 
rate would be $5 \times 10^{-8}$ to $3 \times 10^{-7} M_\sun$~yr$^{-1}$ \citep{sic08}.
Similarly, measuring also the H$_\alpha$ velocity, \citet{sem15} derived 
$1.8 \times 10^{-7}M_\sun$~yr$^{-1}$.  \citet{gia18} presented spectra of GM\,Cep at different 
brightness phases and, on the basis of the dereddened H$_\alpha$ luminosity and its relation to 
the accretion luminosity \citep{alc17}, and then to the accretion rate \citep{gul98}, derived an average
accretion rate of $3.5 \times 10^{-8} M_\sun$~yr$^{-1}$ with no significant temporal variations. 
Each of these methods has its limitation.  The $U$-band flux may be contributed by thermal emission 
from the hot boundary layer (the accretion funnel) between the star and the disk.  
The H$_\alpha$ emission, on the other hand, may be contaminated by absorption in the H$_\alpha$ profile, 
or by chromospheric contribution not related to accretion. 
In any case, GM\,Cep does not seem to be unusually active in accretion activity compared to typical T Taur stars 
or Herbig Ae/Be stars.  The prominent flux variations are the consequences of dust extinction, 
not the FUor kind of flares.  
In Figure~\ref{fig:lc}(a) and (b), the epoches at which literature spectroscopic measurements are available 
are marked, at date 39091 \citep{sic08} and at date 41645 \citep{sem15}, both when the star 
was in a bright state, and at date $\sim45080$ \citep{gia18} when the star was in a faint state.  Among the three 
datasets, the accretion rate does not seem to correlate with the apparent brightness.

\subsubsection{Event Duration and Extinction }

\begin{figure}[htb!]
\centering
 \includegraphics[angle=0,width=\textwidth]{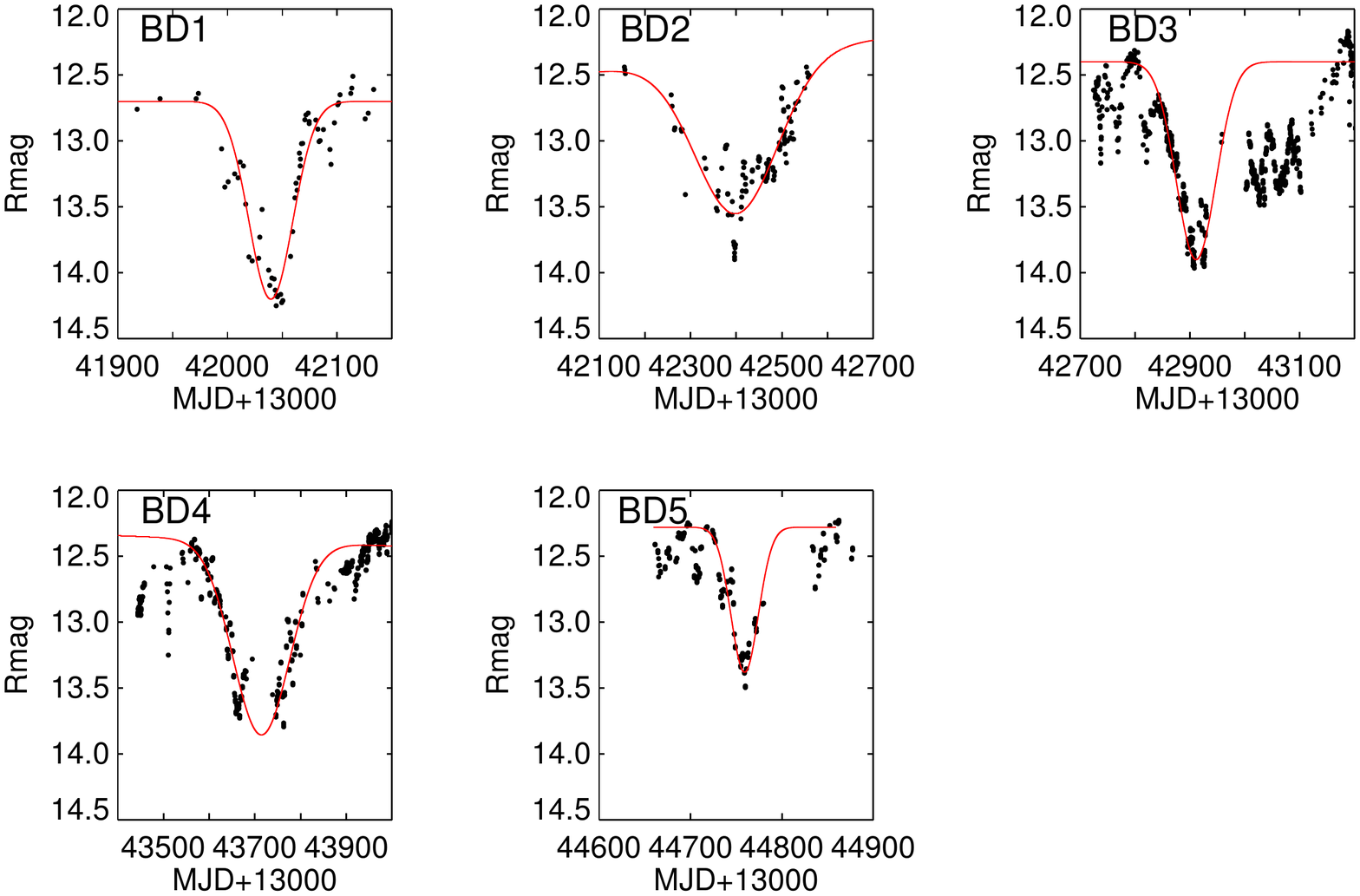}
\caption{The Gaussian fitting to each of the major flux drop events.  
	}
\label{fig:gau}
\end{figure}

We parameterize a flux drop event by its duration and the maximum depth, with a least-squares
fit by a Gaussian function.  Only events sampled at more than half of the duration, e.g., an event lasting 
for roughly 10 days must have been observed for more than 5 nights, are considered to have sufficient 
temporal coverage to be included in the analysis.  Figure~\ref{fig:gau} illustrates how 
the duration, taken as five times the standard deviation, or about 5\% below the continuum, and the 
depth, as the minimum of the Gaussian function, are derived for each major event. 
The parameters are summarized in Table~\ref{tab:duraDep}, in which the columns list for 
each event the identification, the MJD, duration, depths in $B$-, $V$- and $R$-bands, 
and the comments.  

\begin{figure}[htb!]
\centering
 \includegraphics[angle=0,width=\textwidth]{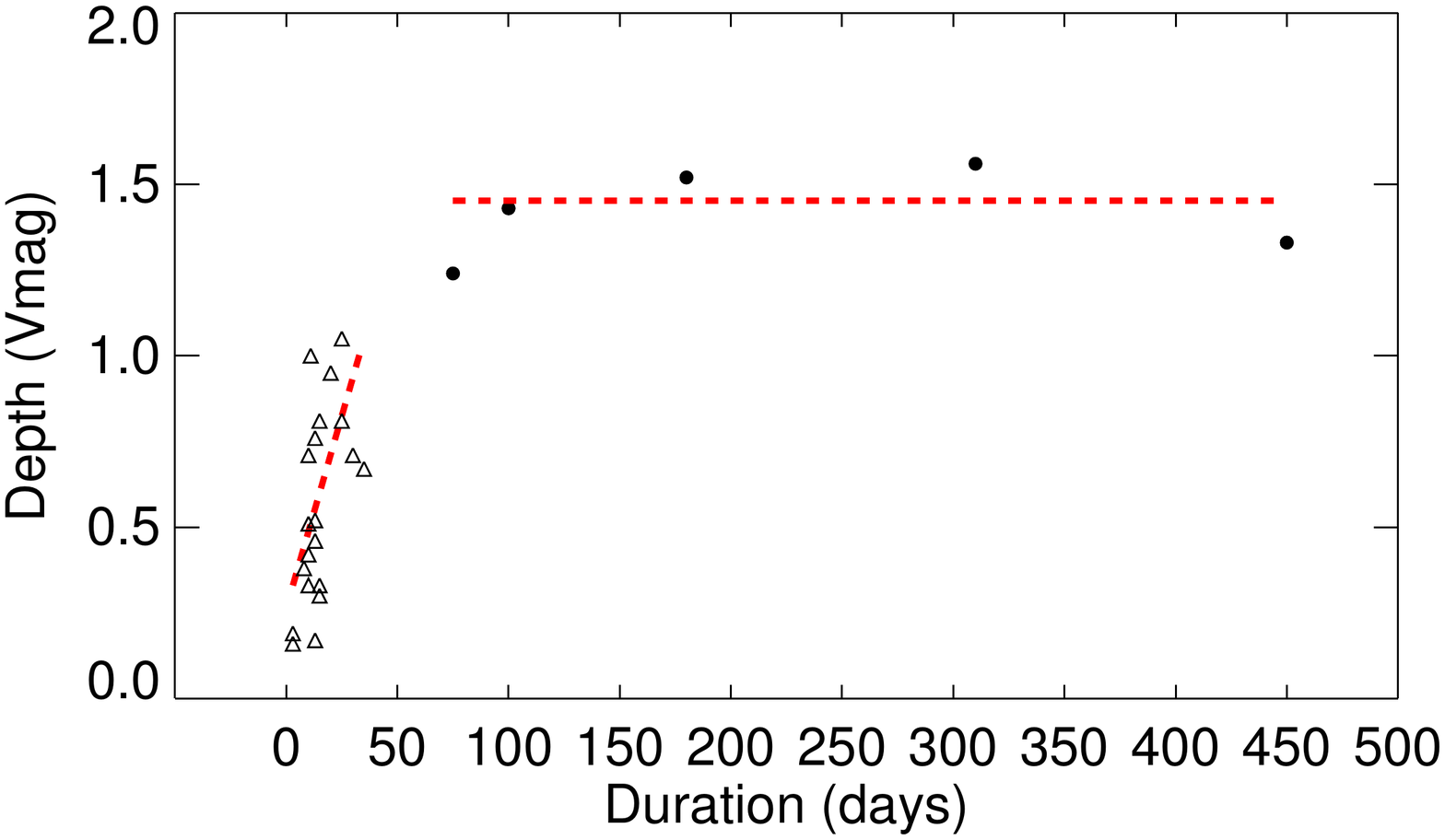}
\caption{Depth versus duration of occultation events.  Each event is parameterized 
	by a Gaussian fit to the light curve as illustrated in Figure~\ref{fig:gau}.  
	There is a linear trend for short events (triangles), whereas for long events (circles)
	the extinction depth levels off.  } 
\label{fig:duraDep}
\end{figure}

Figure~\ref{fig:duraDep} exhibits the duration versus depth of the flux drop events.  
Two distinct classes of events emerge.  For the short events the duration in general 
lengthens with the depth, roughly amounting to $A_V\sim1$~mag per 30 days.  This is understood as 
the various sizes of occulting clumps, so a larger clump leads to a longer event along 
with a deeper minimum.  The extinction depth levels off for longer ($\gtrsim100$ days) events 
to $A_V\sim1.5$~mag, suggesting that these events are not caused by ever larger clumps.  
We propose that each long event consists of a series of events, or a continuous event, 
by clumps distributed along a string or a spiral arm.  In this case, the duration gets longer, but the 
depth is not deeper.    

The depth-duration relation of T Tauri stars has been discussed by \citet{fin13} with 3 yr 
monitoring of Palomar Transient Factory (PTF) for the North America Nebula complex. In their 
sample of 29 stars, there are fading events with a variety of depth (up to $\sim2$~mag) and duration 
(1--100~days).  \citet{sta15}, with a high-cadence light curve from the $CoRoT$ campaign for NGC\,2264, identified 
YSO fading events up to 1~mag.  \citet{guo18} summarized event parameters for different stars, including those 
in \citet{sta15}, and found those with durations less than 10~days varied typically with a depth of $\leq1$~mag, 
whereas those lasting more than $\sim20$~days have a roughly constant amplitude ~$\sim2$--3~mag.  All these 
studies made use of samples of different stars with a diverse star/disk masses, ages, inclination angles, 
etc., and no clear correlation was evidenced between depth and duration.  In comparison, our investigation 
is for a single target with distinct correlations for the short and for the long events.

%%%%%%%%%%%%%
%fig5
%%%%%%%%%%%%%%%

%%%%%%%%%%%%%%%%
%fig6
%%%%%%%%%%%%%%%%%%%%%%

%%%%%%%%%%%%%%%
\begin{deluxetable}{llcllll}
\tablecaption{Flux Drop Events. \label{tab:duraDep}		}
\tablehead{
	\colhead{ID} & \colhead{MJD} & \colhead{Duration (days)} & \colhead{ $\Delta B$ (mag)} & 
	\colhead{ $\Delta V$ (mag)} & \colhead{ $\Delta R$ (mag)} &\colhead{(Remarks)}  
	  }
\startdata
\multicolumn{7}{c}{Major Events}  \\  \hline
BD01  & 55039 & 100 & 1.45 & 1.50 & 1.50 &                   \\  
BD02  & 55401 & 450 & 1.45 & 1.40 & 1.20 &                   \\
BD03  & 55910 & 180 & 1.70 & 1.60 & 1.50 &                   \\
BD04  & 56713 & 310 & 1.75 & 1.64 & 1.47 &                   \\
%BD5  &   -   & 120 & 1.10 & 1.10 & 0.90 & not gaussian\\
BD05  & 57759 & 75  & 1.45 & 1.30 & 1.10 &                   \\ 
\hline
\multicolumn{7}{c}{Minor Events}  \\  \hline
SD01  & 55736 & 10 & 0.79 & 0.75 & 0.67 &                  \\
SD02  & 55767 & 30 & 0.82 & 0.75 & 0.63 &                  \\
SD03  & 55818 & 35 & 0.80 & 0.70 & 0.65 &                  \\
SD04  & 56205 & 25 & 1.05 & 0.85 & 0.78 &                  \\
SD05  & 56415 & 13 & 0.87 & 0.80 & 0.72 &                  \\
SD06  & 56429 & 10 & 0.52 & 0.44 & 0.40 &                  \\
SD07  & 56510 & 11 & 0.65 & 1.05 & 0.70 & $V$ includes AAVSO data\\
SD08  & 56553 & 15 & 0.37 & 0.32 & 0.30 &                  \\
SD09  & 56763 & 13 & 0.35 & 0.55 & 0.68 &                  \\
SD10 & 56784 & 13 & 0.55 & 0.48 & 0.48 &                  \\
SD11 & 56865 & 13 &  -   & 0.40 &  -   &                  \\
SD12 & 56944 & 13 & 0.33 & 0.18 & 0.20 &                  \\
SD13 & 56972 & 3  & 0.22 & 0.17 & 0.15 &                  \\
SD14 & 56989 & 3  & 0.22 & 0.20 & 0.18 &                  \\
SD15 & 57184 & 8  & 0.49 & 0.40 & 0.36 &                  \\
SD16 & 57263 & 25 & 1.10 & 1.10 & 1.40 & incomplete sampling in $B$ and $V$\\
SD17 & 57291 & 10 & 0.40 & 0.35 & 0.30 &                  \\
SD18 & 57333 & 10 & 0.30 & 0.35 & 0.35 &                  \\
SD19 & 57415 & 28 & 0.95 &  -   & 0.87 &                  \\
SD20 & 57511 & 20 & 1.10 & 1.00 & 0.92 &                  \\
SD21 & 57591 & 15 & 0.45 & 0.35 & 0.30 &                  \\
SD22 & 57656 & 10 & 0.61 & 0.54 & 0.48 &                  \\
SD23 & 57946 & 15 & 1.05 & 0.85 & 0.80 &                  \\
\enddata
\end{deluxetable} 
%%%%%%%%%%%%%%%%%%%%

\subsection{Color Variations}  \label{sec:color}

\begin{figure}[htb!]
\centering
 \includegraphics[angle=0, height=0.45\textheight]{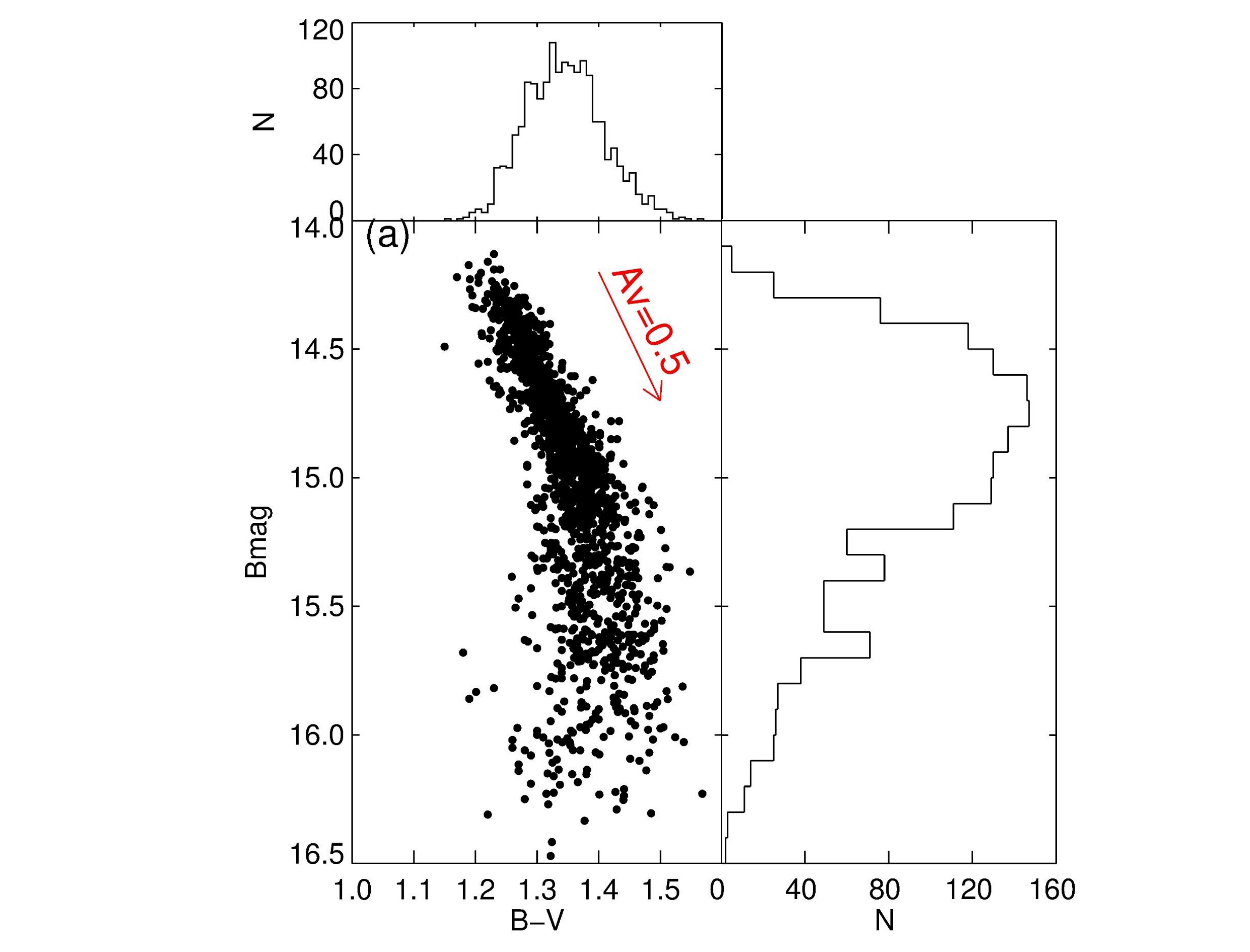}
 \includegraphics[angle=0, height=0.45\textheight]{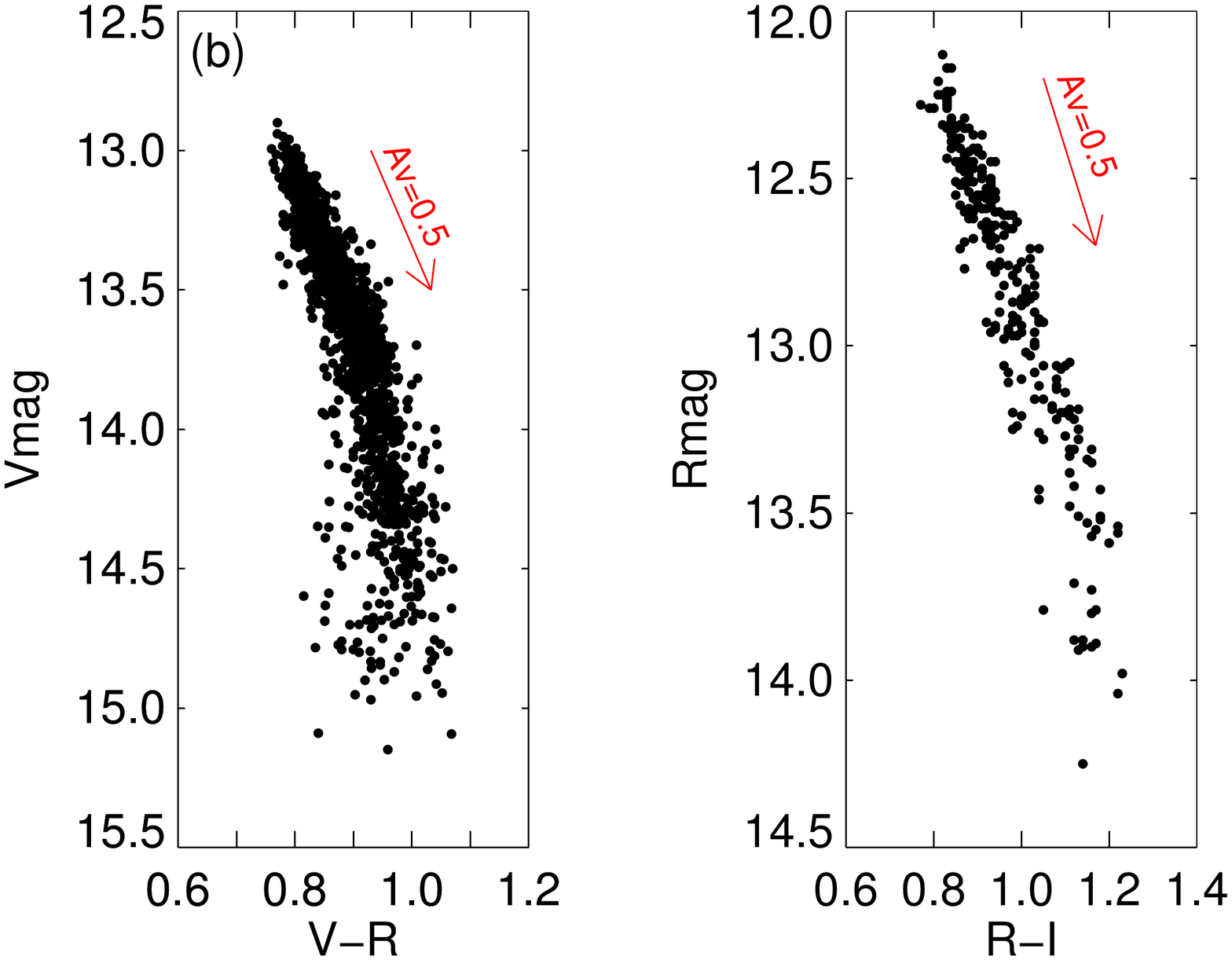}
	\caption{(a) The $B$ magnitude versus $B-V$ color for GM\,Cep, using data in Figure~
	\ref{fig:lc}. The panel on the right plots the histogram of the brightness in $B$, 
	whereas the panel on the top plots the histogram of the $B-V$ color.  The arrow marks 
	the reddening vector for $A_V=0.5$~mag assuming a total-to-selective extinction of 
	$R_V=5.0$. 
	(b) The same as in (a) but for $V$ versus $V-R$ and $R$ versus $R-I$. }
\label{fig:cmd}
\end{figure}

Along with the light curves, Figure~\ref{fig:lc} also presents the $B-V$ color curve, i.e., 
the temporal variation.  Figure~\ref{fig:cmd}a illustrates how the $B$ magnitude of GM\,Cep varies 
with its $B-V$ color.  In this color-magnitude diagram (CMD), GM\,Cep in general 
becomes redder when fainter, suggesting normal interstellar extinction/reddening.  The slope of 
the reddening vector, marked by an arrow, is consistent with a total-to-selective extinction law 
of $R_V=5$ \citep{mat90}, rather than with the nominal $R_V=3$, implying larger dust grains than in the 
diffuse interstellar clouds.  Between $B\sim15.2$~mag and $B\sim15.7$~mag, the extinction appears 
independent of the ($B-V$) color, indicative of gray extinction by even larger grains ($> 10~\micron$, 
\citep{eir02}.  The trend is yet different toward the faint state; namely the color turns bluer when fainter.  
This color reversal, or the ``bluing effect'', 
has been known \citep{bib90,gri94,gra95,her99,sem15}, with the widely accepted explanation being 
that during the flux minimum, when direct star light is heavily obscured by circumstellar dust, 
the emerging light is dominated by forward scattered radiation into the field of view.  

The bluing phenomenon is also illustrated in Figure~\ref{fig:lc}, where a few deep minima are marked, 
each by a thick red line, during which the corresponding color turns blue near the flux minimum.  
Additional CMDs in $V$ versus $V-R$, and $R$ versus $R-I$, where the data in 
$I$ are adopted from those reported by \citet{sem15}, indicate also normal reddening in the 
bright state, whereas the bluing tends to subside toward longer wavelengths, in support of the scattering 
origin, as shown in Figure~\ref{fig:cmd}b.

%%%%%%%%%%%%%%%
%fig7
%%%%%%%%%%%%%%%%%%

\subsection{Polarization }

\begin{figure}[htb!]
\centering
 \includegraphics[angle=270, width=1\textwidth]{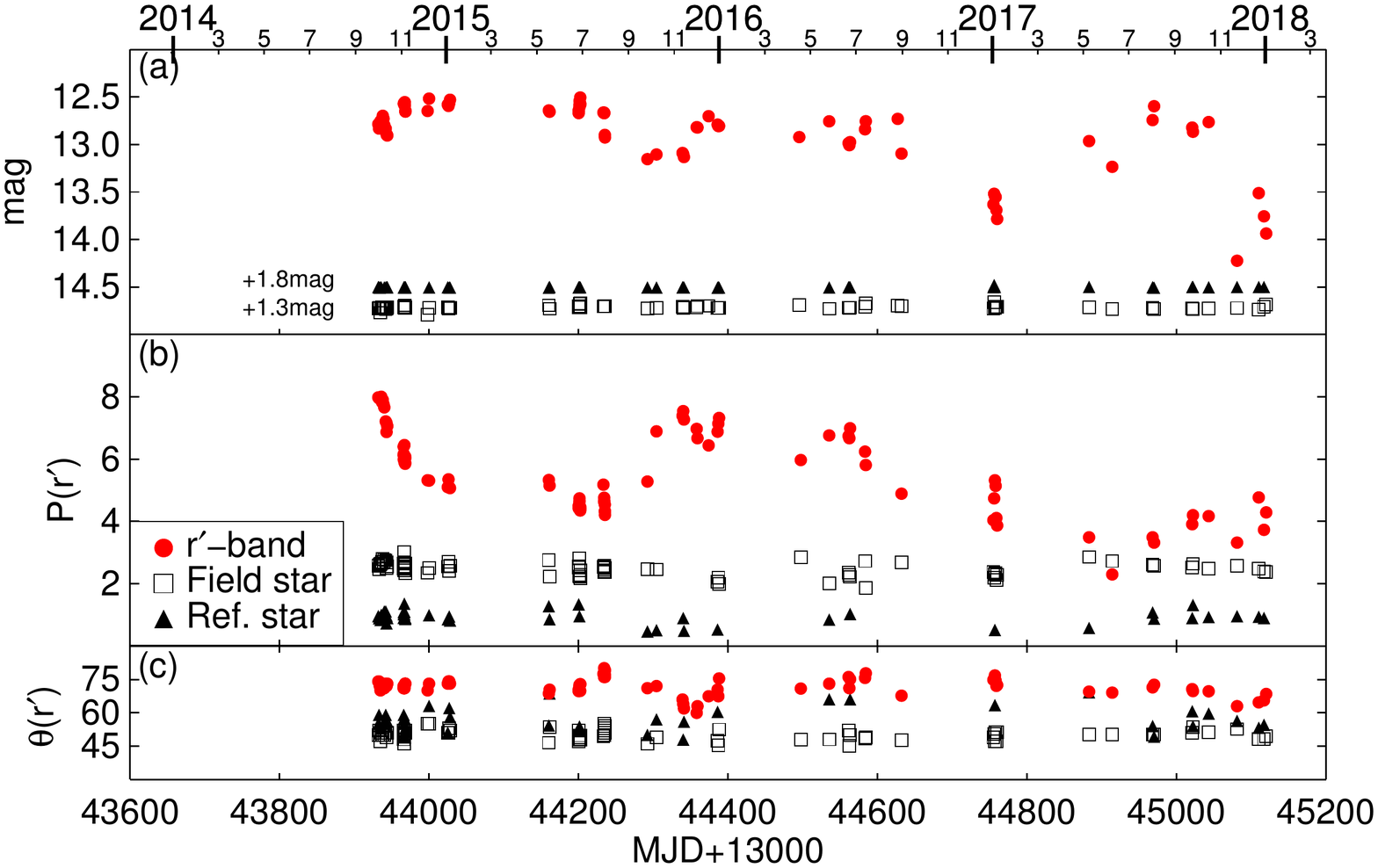}
	\caption{(a)~The $r^{\prime}$-band light curve (in black) for GM\,Cep, together with one of the photometric reference stars (filled triangles) and one 
	field star (squares), in the same field of images. 
	(b)~The changing polarization level of GM\,Cep, in comparison to the two comparison stars. 
	(c)~The polarization angle for GM\,Cep remaining steady (72$\degr$) during three 
	years of monitoring.    }
\label{fig:rpol}
\end{figure}

Figure~\ref{fig:rpol} presents the linear polarization in $r^{\prime}$-band of GM\,Cep, and of 
two comparison stars including one of the photometric reference stars and a field star.  
GM\,Cep displays a varying polarization with $P=3\%$--8\% but with an almost constant 
position angle of $\sim72\degr$.  The two comparison stars remain steadily polarized, each of 
$P \lesssim2\%$ with a variation $\lesssim1$\%.  

Adding up the TRIPOL measurements at four polarizer angles gives the total flux.  
As seen in Figure~\ref{fig:rpol}, the TRIPOL $r^{\prime}$ light curve, albeit with lower cadence, 
allows for diagnosis of simultaneous photometric and polarimetric behavior. The broadband light 
curves in turn serve to indicate the overall brightness states at which the polarization data are taken. 

\begin{figure}[h]
\centering
\includegraphics[width=0.9\textwidth]{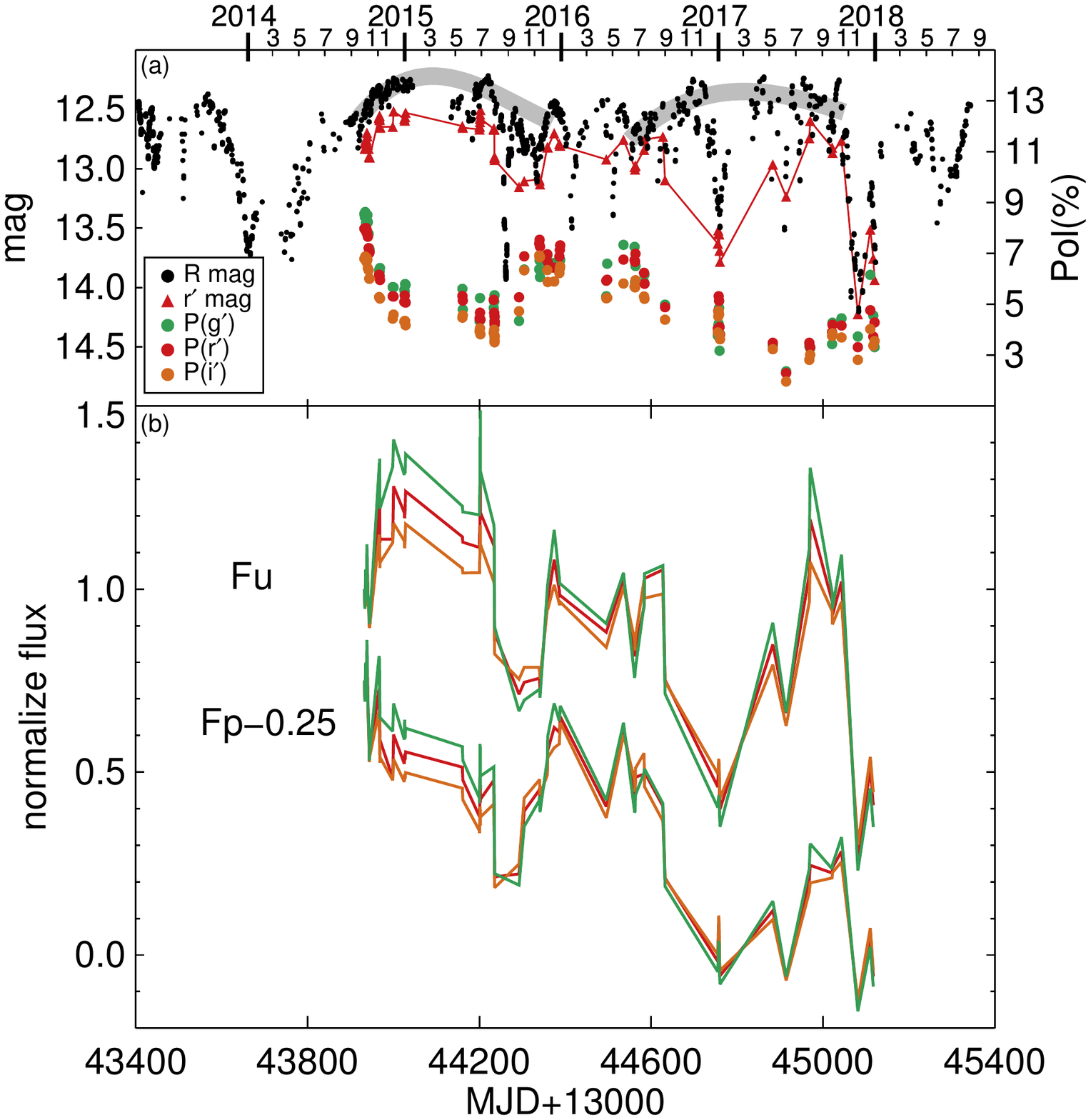}
	\caption{
	(a) The photopolarimetric $r^{\prime}$-band light curve (in red) 
	vs. the $R$-band light curve (in black), and shown below the polarization levels 
	in $g^{\prime}$ (in green), $r^{\prime}$ (in red), and $i^{\prime}$ (in brown).
	The gray shades represent the slow brightness changes and simultaneous behavior 
	of the polarization.  
	(b) The light curves for unpolarized flux ($F^u$) and polarized flux ($F^p$), with the 
	same color symbols as in (a).}
\label{fig:gripol}
\end{figure}

Figure~\ref{fig:gripol}a plots the polarization in each band, $P_{g^{\prime}}$, $P_{r^{\prime}}$, 
and $P_{i^{\prime}}$.  The polarization exhibits a slowly varying pattern, declining from 6\% to 9\% 
in the fall of 2014 to 3\%--5\% in 2015 July/August, and reclining to 5\% -- 7\% 
near the end of 2015.  A similar pattern seems to exist also in 2017 but with a variation of  
2\% -- 5\%.  
At the same time, the slow brightness change in each case, notwithstanding abrupt flux drops, 
seems to have a reverse trend.  In particular, the smooth brightening in late 2014, 
where polarization data are densely sampled, is clearly associated with a monotonic 
decrease in polarization.  A similar brightness-polarization pattern is seen from early 2017 to 
early 2018, for which the brightening and fading in the light curve is associated with a 
decreasing-turn-increasing trend in polarization.

Note that in general the polarization is higher at shorter wavelengths, but 
at certain epochs, particularly at flux minima, e.g., at the end of 2015 and the 
beginning of 2017, an ``anomalous'' wavelength dependence seems to emerge, so 
that the $g^{\prime}$ band becomes the least polarized.

%%%%%%%%%%%%%%%%%%%%%%%
%fig8
%%%%%%%%%%%%%%%%%%%%%%%%

%%%%%%%%%%%%%%%%%%%%%%%%
%fig9
%%%%%%%%%%%%%%%%%%%%

\section{The Clumpy Disk Structure in GM\,Cep} \label{sec:disk}

The photopolarimetric measurements enable inference on the occultation configuration in a 
qualitative way.  For example, a sequential blockage of the circumstellar environs and the 
star will result in a certain photometric and polarimetric behavior.  
The high-cadence light curves, furthermore, allow quantitative derivation of 
the depth, duration, etc., of the occulting body.  We present the analysis and 
interpretation of both kinds in this section.  

\subsection{Occultation Geometry Inferred By the Polarization Data} 

The level of polarization at wavelength $\lambda$ is defined as 

  \[
    P_\lambda (\%) = \frac{F^p_\lambda}{F^t_\lambda} 
                   = \frac{F^p_\lambda}{F^p_\lambda + F^u_\lambda}
		   = \frac{1}{1+F^u_\lambda/F^p_\lambda}, 
  \]
where $F^t$ is the total flux, which is decomposed into polarized flux ($F^p$) 
and unpolarized flux ($F^u$), with $F^t=F^p + F^u$.   
At each observing epoch, $P_\lambda$ and $F^t$ are measured, therefore $F^p$ and $F^u$ 
can be derived.  In general the starlight is not polarized, but the scattered light 
from the inner gaseous envelope/disk is, which is fainter and bluer in color than the direct 
starlight.  

The temporal variations of $F^p_\lambda$, $F^u_\lambda$, and $F^t_\lambda$, plus the 
wavelength dependence of these variations, provide clues on the geometry of a clump, or 
a string of clumps, relative to the stellar system (star plus disk). 
The last part of the equation suggests that (1) if $F^u_\lambda$ remains the same,
$P_\lambda$ changes with $F^p_\lambda$ in the sense that as 
$F^p_\lambda$ decreases, so does $P_\lambda$.  The dust reddening by occultation 
makes this dependence stronger at shorter wavelengths.  But (2) if $F^u_\lambda$ 
changes, because it dominates the brightness over $F^p_\lambda$, 
so, for example, as $F^u_\lambda$ decreases, $P_\lambda$ increases.  

Figure~\ref{fig:gripol}b exhibits how the decomposed 
polarized ($F^p$) and unpolarized ($F^u$) components 
vary, respectively, at different wavelengths.  To facilitate the comparison, each curve is scaled to 
its first data point to demonstrate the relative level of flux changes.  The decomposition 
makes it clear that the decreasing polarization near the end of 2014, 
with $P_{g^{\prime}} > P_{r^{\prime}} > P_{i^{\prime}}$ (see Fig.~\ref{fig:gripol}a), 
corresponding to the brightening of the star system, is the result of a fading $F^p_\lambda$ 
alongside with a brightening $F^u_\lambda$, as evidenced in Fig.~\ref{fig:gripol}b, 
both leading to a decreasing $P_\lambda$ in every wavelength.  
In the occultation scenario, the star system would be 
just coming out of a major event, and during such an egress, the clump was 
unveiling the star and blocking a progressively larger part of the envelope.  
Incidentally the deep flux drop event at the 
beginning of 2017 has polarization measured.  At the brightness minimum, the level of 
polarization changes little, but with the anomaly $P_{r^{\prime}} > P_{i^{\prime}} > P_{g^{\prime}}$.  
Inspection of the decomposition result reveals that both $F^u_\lambda$ and $F^p_\lambda$ decline 
to almost an all-time low, particularly at shorter wavelengths.  This is the configuration 
when the star and the envelope are both heavily obscured.

On YSO photometric and polarimetric variability, \citet{woo96} and \citet{sta99} modeled   
the rotationally modulated multiwavelength photopolarization due to scattering of light by stellar hot spots, 
under different simulation parameters, such as the size and latitude of the hot spot, inclination, 
truncation radius, and geometry (e.g., flat or flared) of the disk.  In general, the simulations suggested 
an amplitude of polarization variability less than about $1\%$. 
The polarization variability due to a warped disk is similarly low, as demonstrated in the case
of AA\,Tau, a prototype of dippers, with a variation of $\sim0.5\%$ in the $V$-band during the occultation 
\citet{osu05}.  

Recent modeling by \citet{kes16} of the photopolarimetric variability of YSOs plus accretion disks 
considered the spot temperature, radius of inner disk, structure, and inclination 
of the warp disk.  Only star and dust emission was included, with no gas emission, but still, 
the typical polarization is expected to vary by less than $\sim1\%$. It is interesting that the  
polarization level of $I$-band normally is always higher than that of the $V$-band, 
consistent with the wavelength dependence of our observations, albeit with limited 
time coverage, near flux minima.  Hot starspots or a warped inner disk alone apparently cannot 
account for the large polarization variability seen in GM\,Cep.  An additional gaseous 
envelope likely plays an important role.

\subsection{Clump Parameters By the Light-curve Analysis} 

The long-term light curves render conclusive evidence that the major flux drops detected in 
GM\,Cep are caused by occultation of the young star and the envelope by circumstellar dust clumps. 
These dust grains are large in size, inferred by the reddening law (see \S\ref{sec:color}), 
and distributed in a highly nonuniform manner.  This density inhomogeneity could signify the 
protoplanetary disk evolution in transition from grain growth (of \micron\ size) to planetesimal 
formation (of kilometer size) \citep{che12}.  

Accretion plus viscous dissipation heats up a young stellar disk early on.  As the accretion 
subsides and grains get clumpy, the disk becomes passive, in the sense that the dust absorbs 
starlight, warms up, and reradiates in infrared \citep{chi97}.  
The frequent occultation events imply a geometry that would have led to a significant stellar extinction 
and a flat spectral energy distribution (SED).
Instead, however, because of the grain coagulation, 
GM\,Cep (1) has a moderate $A_V=2$--3~mag, partly of interstellar origin, despite the copious dust 
content evidenced by the elevated fluxes in far-infrared and submillimeter wavelengths \citep{sic08}, 
and also (2) has an SED characteristic of a T~Tauri star \citep{sic08} with a noticeable 
infrared excess.  In a passive disk, hydrostatic equilibrium results in 
a structure to flare outward \citep{key87,chi97}, so the dust intercepts more starlight 
than a geometrically thin disk.  
  
Ring- or spiral-like structure in YSO disks seems ubiquitous, as evidenced by, e.g., 
recent ALMA imaging in molecular lines or in continuum of the Herbig Ae/Be 
star AB\,Aur \citep{tan12,tan17}, the class II object Elias\,2-27 \citep{per16}, 
or by HiCIAO/Subaru polarimetric imaging of FUors \citep{liu16}.  Such a structure 
may be induced by a planet companion \citep{zhu15} or by gravitational instability \citep{kra16}.   
All these rings or spirals have some tens to hundreds of astronomical units in extents. 

The most enlightening finding relevant to our work is the detection in the T Tauri star HL\,Tau 
at 7~mm of a distribution of clumps along the main ring of thermalized dust found earlier 
by at shorter wavelengths, where large grains reside \cite[][see their Figure~2]{car16}.  
The most prominent one, at $\sim0\farcs1$ from the star, or $\sim14$~au at a distance 
of 140~pc, with an estimated mass of 3--8~$M_\earth$, is considered by these authors 
as a possible planetary embryo.  

We have no knowledge of the location of the (strings of) clumps in the GM\,Cep 
disk, or of their geometric shape.  But we present the following exercise, using theoretical 
disk models, to shed light on the possible constraints on clump parameters.    
The largest clumps in GM\,Cep, as seen in Figure~\ref{fig:duraDep}, cause a maximal 
extinction of $A^c_V=1.5$~mag with a time scale of $\sim50$~days. 
Note that here $A^c_V$ refers to the extinction caused by the occultation of the clump, to 
be distinguished from the interstellar plus circumstellar extinction of the star.  
The maximal extinction provides information on the column density of dust, and the 
duration time on the scale of the clump.  
The fiducial disk by \citet{chi97} adopts a stellar temperature $T_{*}=4000$~K, mass 
$M_*=0.5~M_\sun$, and radius $R_*=2.5~R_\sun$.  With veiling and line blending due to fast rotation, 
the spectral type of GM\,Cep is uncertain, ranging from an F9 \citep{hua13} to G5/K3 
\citep{sic08}.  In any case the star is hotter (with higher pressure) but more massive (with 
stronger gravitational pull), and the hydrostatic conditions in the disk turn out to be similar.  
This means the disk height ($H$) is scaled with the radius 
($r$) $H/r \approx 0.17 (r/{\rm au})^{2/7}$ \citep{chi97}.  
A clump at $r=14$~au thus would subtend an opening angle (viewing the rim from the star) 
of $\sim20\degr$; at $r=1$~au, the 
angle would become $\sim 10\degr$, for which the disk has to be close to edge-on for occultation 
to take place.  Assuming 2~$M_\sun$ for GM\,Cep, a clump at 4--14~au has a 
projected Keplerian speed up to 11~km~s$^{-1}$.  So for a clump to traverse the GM\,Cep system, the 
linear size would be 0.3~au for $r=14$~au.  In the case $r=1$~au, the orbital speed 
is faster, so the linear scale would be 1.2~au.  

Alternatively, the clumps may be located closer in to the central star.  
The disk may not be monotonically flared, as the innermost disk is irradiated by starlight, and  
dust evaporation at temperature $T_{\rm evap}\sim 1500$~K results in an inner hole, hence an inner 
rim or ``wall'' in the flaring disk, which accounts for the bump near 2--3~\micron\ 
observed in the SEDs of some YSOs \citep{dul01,eis04}.  This temperature corresponds to 
a distance from the central star, 
$r_{\rm rim}=(L_*/4\pi T^4_{\rm rim}\sigma)^{1/2} ( 1+ (H_{\rm rim}/r_{\rm rim})) ^{1/2}$, 
where $L_*$ is the luminosity of the star, $T_{\rm rim}=T_{\rm evap}$ is the temperature 
at the rim, $H_{\rm rim}$ is the vertical height of the inner rim, and $\sigma$ is 
the Stefan--Boltzmann constant \citep{dul01}.  Given $L_*=26~L_\sun$ for GM\,Cep \citep{sic08}, 
adopting $H_{\rm rim}/r_{\rm rim}=0.2$ \citep{dul01}, the estimated inner rim radius is 
roughly $r_{\rm rim} \sim 0.4$~au, corresponding to an opening 
angle $\arctan{(H_{\rm rim}/r_{\rm rim})} \sim11\degr$.  Even 
though the chance of occultation is higher with a clump closer to the star, a faster Keplerian 
speed would lead to a linear size of 1.7~au.  
We conclude that the ``clump,'' or the region of density enhancement in the disk 
has a length scale up to roughly 0.1--1~au cross the line of sight.

The depth, or the length scale along the line of sight, is related to the maximum $A^c_V=1.5$~mag, 
or the column density of dust.  Integration requires detailed disk structure, such as the 
vertical and radial density profiles, grain size distribution, midplane settling, etc.  
Such a complexity is beyond the scope of this paper and in fact not justified by our 
data.  Here we again attempt to gain some physical insights on the clump properties.  

For a uniform disk, the volume mass density of dust 
$m_d = (N_d / \ell) \, M_{\rm grain}$, where $N_d$ is the column density of dust, $\ell$ 
is the length of the sightline through the dusty medium, and $M_{\rm grain}$ is the 
mass of each grain.  Each term is evaluated as follows. 

The column density $N_d$ is related to the extinction: $A^c_V=1.086 \tau_V = N_d \sigma_d Q_{\rm ext}$, 
where $\tau_V$ is the optical depth at $V$-band, $\sigma_d = \pi a^2$ is the geometric cross 
section of each (assuming spherical) grain of radius $a$, and $Q_{\rm ext}$ is the optical 
extinction coefficient, which, for grains large in size compared to the 
wavelength ($2 \pi a >> \lambda$), $Q_{\rm ext} \approx 2$ \citep{spi78,van57}.
Therefore, 
$N_d=1.6 \times 10^5\, A^c_V\, [10~\micron/a]^2$~cm$^{-2}$, and for each dust grain, 
assuming a material bulk density of 2~g~cm$^{-3}$, the mass is 
$M_{\rm grain}=8.4 \times 10^{-9} [a/10\micron]^3$~g.  Given a gas density $n_g$, 
and a nominal gas-to-dust mass ratio of 100, $m_d=n_g m_H /100$, and so

  \[
	  \ell = \frac{5.4 \times 10^{9}}{n_g} \, A_V \, (\frac{a}{10~\micron})~~~{\rm [au]}.
  \]

For GM\,Cep, $A^c_V=1.5$~mag, and adopting a gas density $n_g=10^{10}$ \citep{bar05}, 
$\ell\sim 0.8$~au for $a=10$~\micron\ grains.  For truly large grains, such as $a=1$~mm, 
the extinction efficiency becomes much smaller, thus $\ell$ 100 times longer, 
to $\ell\sim 80$~au.  

Admittedly, none of the simple assumptions we have made in the estimation is likely 
valid.  Still, it is assuring that both the crossing time and the flux drop of occultation 
by a dust clump could end up with reasonable solutions, namely a region tens of astronomical units across 
in the young stellar disk, perhaps in a ring or a spiral configuration located tens of astronomical units 
from the star, consisting of primarily 10~\micron\ grains or larger.  
Given the overall low extinction of the star, small grains likely exist but not in quantity, 
as they had been agglomerated into large bodies.

%Star formation in Tr\,37 may not be coeval.  In addition to the majority of members of ages about 
%4~Myr, which GM\,Cep (given its spatial location) should belong to \citep{sic08}, there exists a 
%younger population of ages about 1~Myr to the west near the bright-rimmed globule \citep{sic05}.  
%An older age favors the grain growth process to 10~\micron\ or larger.  GM\,Cep, with a spectral 
%type F or G, therefore serves 
%as an example of disk evolution leading to planetesimal formation, linking that in typical T Tauri 
%stars (mostly K to M types) to that in Herbig Ae/Be stars (as late as F types).

\section{Conclusion} \label{sec:conclusion}

Optical photometric and polarimetric monitoring of the UX Ori star GM\,Cep for nearly a decade reveals 
variations in brightness and in polarization of different amplitude and time scales.   
The essential results of our study are:

\begin{itemize} 

\item GM\,Cep exhibits (1) brightness fluctuations $\lesssim0.05$~mag on time scales of 
	days, due partly to rotational modulation by surface starspots with a period 
	of 3.43~days, and partly to accretion activity; (2) minor flux drops of amplitude 0.2--1.0~mag with 
	duration of days to weeks; and (3) major flux drops up to 2.5~mag, each lasting 
	for months, with a recurrent time, but not exactly periodic, of about 2 years.

\item The flux drops arise from occultation of the star and gaseous envelope by orbiting 
	dust clumps of various sizes.   

\item The star experiences normal dust reddening by large grains, i.e., 
	the star becomes redder when fainter, except at the brightness minimum 
	during which the star turns bluer when fainter.  

\item The maximum depth of an occultation event is proportional to the duration, 
	about 1~mag per 30 days, for the events lasting less than $\sim 50$ days, 
	a result of occultation by clumps of varying sizes.  
	For the events longer than about 100 days, the maximum depth is independent of 
	the duration and remains $A_V\sim1.5$~mag, a consequence of transiting 
	strings or layers of clumps.  

\item The $g^{\prime} r^{\prime} i^{\prime}$ polarization levels change between 3\% and 8\%, and vary 
	inversely with the slow brightness change, while the polarization angle remains constant.  
	The polarization is generally higher at shorter wavelengths, but at flux minima, there 
	is a reversal of wavelength dependence, e.g., the $g^{\prime}$-band becomes the least 
	polarized.  Temporal variations of polarization versus brightness, once the total 
	light is decomposed into polarized and unpolarized components, allow diagnosis of 
	the occultation circumstances of the dust clumps relative to the star and envelope.    

\item Our data do not provide direct information on the size or location of the clumps, but 
	the duration of an occultation sets constraints on the transverse size scale of the 
	clump, while the maximum extinction depth is a measure of the column density of dust, 
	hence a dependence of the line-of-sight length through the dusty medium.  It is possible 
	that GM\,Cep is an edge-on manifestation of the ring- or spiral-like structures found recently 
	in young stars with imaging in infrared of scattered light, or in submillimeter of 
	dust emission.
	
\end{itemize}

\acknowledgments

The NCU group acknowledges the financial support of the grants 
MOST 106-2112-M-008-005-MY3 and MOST 105-2119-M-008-028-MY3. 
We greatly thank the Jena group H. Gilbert, T. Zehe, T. Heyne, A. Pannicke, and 
C. Marka for kindly providing us with their efforts on acquiring data from Jena Observatory, which is 
operated by the Astrophysical Institute of the Friedrich-Schiller-University. Furthermore, we 
would like to thank the Thuringian State (Th\"uringer Ministerium f\"ur Bildung, Wissenschaft 
und Kultur) in project number B~515-07010 for financial support.
The work by the Xinjiang Observatory group was in part supported by the program of 
the Light in China's Western Region (LCWR, grant No.~2015-XBQN-A-02) and National Natural 
Science Foundation of China (grant No.~11661161016). J.~Budaj, Z.~Garai, and T.~Pribulla acknowledge 
VEGA 2/0031/18 and APVV 15-0458 grants as well as V. Kollar, J. Lopatovsky, N. Shagatova, S. Shugarov,
and R. Komzik for their help with some observations.
This research has made use of the International Variable Star Index (VSX) database, operated at AAVSO, 
Cambridge, Massachusetts, USA.
We thank the referee for constructive comments that greatly improve the quality of the paper.

\end{document}